\documentclass[review]{fcs}
\definecolor{d1}{HTML}{9B9C2D}
\definecolor{d2}{HTML}{D8DEB8}
\definecolor{m1}{HTML}{2AA371}
\definecolor{m2}{HTML}{B7E3CF}
\definecolor{s1}{HTML}{DB726B}
\definecolor{s2}{HTML}{E8A7A2}
\definecolor{v1}{HTML}{A977A6}
\definecolor{v2}{HTML}{D4BAD2}
\definecolor{h1}{HTML}{2B9CD7}
\definecolor{h2}{HTML}{91CDEB}

\title{Human-Computer Interaction and Visualization in Natural Language Generation Models: Applications, Challenges, and Opportunities}
\author[1]{Yunchao Wang}
\author*[1,2]{Guodao Sun}
\author[1]{Zihang Fu}
\author[1]{Ronghua Liang}
\address[1]{College of Computer Science and Technology, Zhejiang University of Technology, Hangzhou 310023, China}
\address[2]{Zhejiang Key Laboratory of Visual Information Intelligent Processing, Hangzhou 310023, China}
\fcssetup{
  received       = {month dd, yyyy},
  accepted       = {month dd, yyyy},
  corr-email     = {guodao@zjut.edu.cn},
}
\begin{abstract}
Natural language generation (NLG) models have emerged as a focal point of research within natural language processing (NLP), exhibiting remarkable performance in tasks such as text composition and dialogue generation. However, their intricate architectures and extensive model parameters pose significant challenges to interpretability, limiting their applicability in high-stakes decision-making scenarios. To address this issue, human-computer interaction (HCI) and visualization techniques offer promising avenues to enhance the transparency and usability of NLG models by making their decision-making processes more interpretable. In this paper, we provide a comprehensive investigation into the roles, limitations, and impact of HCI and visualization in facilitating human understanding and control over NLG systems. We introduce a taxonomy of interaction methods and visualization techniques, categorizing three major research domains and their corresponding six key tasks in the application of NLG models. Finally, we summarize the shortcomings in the existing work and investigate the key challenges and emerging opportunities in the era of large language models (LLMs).
\footnote{The article has been accepted by \textit{Frontiers of Computer Science (FCS)}, with the DOI: 10.1007/s11704-025-50356-6.}
\end{abstract}

\begin{document}

\section{Introduction}
NLG models have emerged as a focal point in the field of NLP. Many technology companies have introduced NLG-powered products that enable users to obtain desired responses at minimal cost~\cite{petridis2023promptinfuser}. Currently, NLG models have been widely adopted in various applications, including code generation~\cite{li2021nbsearch,tang2024mavidsql}, text composition~\cite{peng2020exploring,chakrabarty2022help,petridis2023anglekindling}, image captioning~\cite{mao2016generation,hsu2019visual,stefanini2022show,padmakumar2022machine}, question-answering systems~\cite{gao2021meaningful}, and task planning~\cite{shin2023planfitting}. Compared to traditional computational tools, NLG models possess the ability to not only comprehend textual, visual, and video-based semantic information but also generate relevant outputs tailored to user intentions~\cite{wang2019vizseq,singh2023hide}. Against this backdrop, HCI and visualization techniques have garnered increasing attention for their applications in text generation. These technologies offer promising opportunities to enhance user experience and optimize system performance. However, despite the impressive capabilities of NLG models, their complex architectures and massive parameter sizes pose challenges in interpretability~\cite{luo2024local,wu2025visual}. To address this issue, the integration of HCI methods and visualization techniques provides a potential pathway to demystify these ``black-box'' models~\cite{li2023personalized}, thereby improving model transparency, enhancing user experience, and refining system effectiveness.

HCI refers to the communication and interaction between users and computational systems, with the primary objective of designing user-friendly, efficient, and intuitive interaction mechanisms to enhance user experience and system usability~\cite{nguyen2018believe}. In the context of NLG, HCI research focuses on developing optimized interaction interfaces and mechanisms that facilitate seamless text generation, editing, and refinement~\cite{zhou2023recurrentgpt}. This research extends beyond interface design to encompass a comprehensive understanding of user behaviors and needs, enabling the provision of more personalized and intelligent text generation services~\cite{bansal2021does}.

In the domain of NLG, visualization plays a critical role in assisting both text generation and model interpretability. By presenting data intuitively, researchers aim to improve user interaction with complex models~\cite{gehrmann2019gltr}. For example, visualizing the inference process during text generation can enhance model transparency and improve interpretability by making decision-making processes more comprehensible. Additionally, visualization techniques can assist in text proofreading and refinement by providing graphical feedback that helps users identify and correct errors, ultimately improving text quality and accuracy~\cite{hsu2021plot}. Moreover, visualization enables users to gain deeper insights into the model’s decision-making processes and internal structures, making these otherwise complex systems more transparent and accessible.

In this paper, we provide a comprehensive review of recent advancements in HCI methods and visualization techniques in the context of NLG. It analyzes current challenges and unresolved issues while outlining future research directions. By offering an in-depth discussion of these technologies, this review aims to serve as a valuable reference for researchers in related fields and contribute to the further development and application of HCI and visualization techniques in NLG models.

\section{Related Surveys}
In this section, we review several survey papers related to NLG models, which focus on both the model and user perspectives. Papers centered on the model primarily emphasize the ``\textit{model details and principles}''~\cite{rogers2021primer,zhao2024explainability,li2024pre} and ``\textit{model application scenarios}''~\cite{voigt2022and,xu2023multimodal,celikyilmaz2020evaluation}, while those centered on the user focus on the ``\textit{human-in-the-loop workflow and interactivity}''~\cite{wang2021putting,dudley2018review} and the ``\textit{usage patterns across different user groups}''~\cite{zamfirescu2023johnny}. Additionally, researchers have shown considerable interest in the \textit{visualization techniques}~\cite{brath2023role,yang2024foundation,bracsoveanu2020visualizing,hohman2018visual} within NLG models.

Since the introduction of the attention mechanism~\cite{Vaswani2017Attention}, models based on the Transformer architecture have achieved groundbreaking advancements in the field of NLP, with notable examples including BERT~\cite{Devlin2019BERT}, GPT~\cite{brown2020language}, and LLaMA~\cite{touvron2023llama}. Despite their remarkable performance, these models exhibit a high degree of opacity, characteristic of “black-box” systems. This lack of transparency not only impedes model optimization and understanding but also introduces potential risks to downstream applications. Consequently, both experts and non-experts seek to gain deeper insights into the hierarchical structure and underlying principles of these models.
To address this issue, researchers have conducted a systematic investigation into BERTology~\cite{rogers2021primer}, reviewing BERT’s operational mechanisms, information learning, and representation processes. Furthermore, they have summarized key aspects such as BERT’s training objectives, common architectural modifications, issues related to overparameterization, and model compression techniques. Additionally, in the study of interpretability in Transformer-based models, researchers have classified existing methodologies according to training paradigms, distinguishing between the traditional fine-tuning paradigm and the emerging prompting paradigm~\cite{zhao2024explainability}. They have also explored how interpretability techniques can be leveraged to debug and optimize models, ultimately enhancing their overall performance.

Researchers have increasingly recognized the inefficiency of traditional analytical methods, which has hindered both the improvement and widespread adoption of models. Consequently, there is a growing interest in introducing novel techniques to address these limitations. With the emergence of the ``human-in-the-loop'' paradigm, HCI methods have been increasingly integrated into model development. By interacting with real users, language models can continuously receive user feedback and new data, enabling ongoing learning and refinement.
Analyzing the interaction process between users and models facilitates the identification of existing limitations and deficiencies, allowing for targeted adjustments and optimizations to enhance model performance and applicability~\cite{wang2021putting}. As a result, researchers must systematically characterize interactive machine learning from the perspective of interactive systems, identifying key aspects of user interface research~\cite{dudley2018review,lee2024design} and the usage patterns of different demographics~\cite{zamfirescu2023johnny} to unlock more efficient applications of interactive machine learning.

However, existing surveys have not synthesized the HCI and visualization involved in the text generation process based on NLG models. It has been observed that model understanding and application remain focal points and challenges within the field of NLG. Due to its intuitive and effective representation of data, visualization is increasingly used to enhance the interpretability of NLG models. HCI affords users and models the opportunity to evolve collaboratively during the training process, and enables a broader user base to utilize NLG model products. Consequently, this paper sets out from the perspectives of HCI and visualization to delineate the landscape of text generation based on NLG models, aiming to facilitate practitioners in benefiting from a broader array of research outcomes.

\section{Methodology and Taxonomy}
In this section, we will elaborate on the methodology and taxonomy used for collecting and screening academic papers.

\subsection{Methodology}
The primary objective of this survey is to investigate text generation based on NLG models, with a particular focus on technical and instructional papers related to HCI and visualization. This survey systematically examines the entire process from three levels: \textit{Data}, \textit{Model}, and \textit{System}. To delineate the scope of the investigation, four core terms are defined: natural language generation models, text generation, human-computer interaction, and visualization. Finally, this survey emphasizes methods and techniques of HCI and visualization, aiming to identify application directions that have garnered attention in the fields of visualization and HCI.

For research on visualization and HCI, we use the following key terms for literature retrieval: ``text generation'', ``natural language generation models'', ``creative writing'', and ``writing support''. Relevant conferences and journals include but are not limited to \textit{IEEE VIS}, \textit{EuroVis}, \textit{PacificVis}, \textit{CHI}, \textit{UIST}, \textit{IEEE TVCG}, \textit{ACM TOCHI}, and \textit{ACM TIIS}.
Regarding natural language processing and artificial intelligence, our study focuses on core themes such as ``visualization'', ``human-computer collaboration'', and ``writing assistants''. Relevant academic conferences and journals include \textit{NLPCC}, \textit{ACL}, \textit{EMNLP}, \textit{NAACL}, \textit{AAAI}, \textit{TACL}, and \textit{CVMJ}. A preliminary collection has identified a total of 246 papers.
To ensure the scientific rigor and reliability of the literature selection process, we invited three experts with extensive research experience in the fields of visualization, HCI, and NLP to further screen the collected papers. The selection criteria were as follows: (1) the paper must be published in high-impact, widely recognized journals or international academic conferences to ensure its credibility and academic recognition; (2) the content of the paper must be relevant to the three core topics of ``human-computer interaction'', ``visualization'', and ``natural language generation models'' to maintain thematic relevance; (3) the proposed methods or technologies in the paper must demonstrate practical or potential applicability in current or future text generation tasks, ensuring both practical significance and forward-looking value.
The number of papers that met the criteria after screening totaled 78.
The statistics shows that the majority of the collected papers were published between 2016 and 2024.

\begin{figure}[htbp]
\centering
\vspace{-0.6em}
\includegraphics[width=\linewidth]{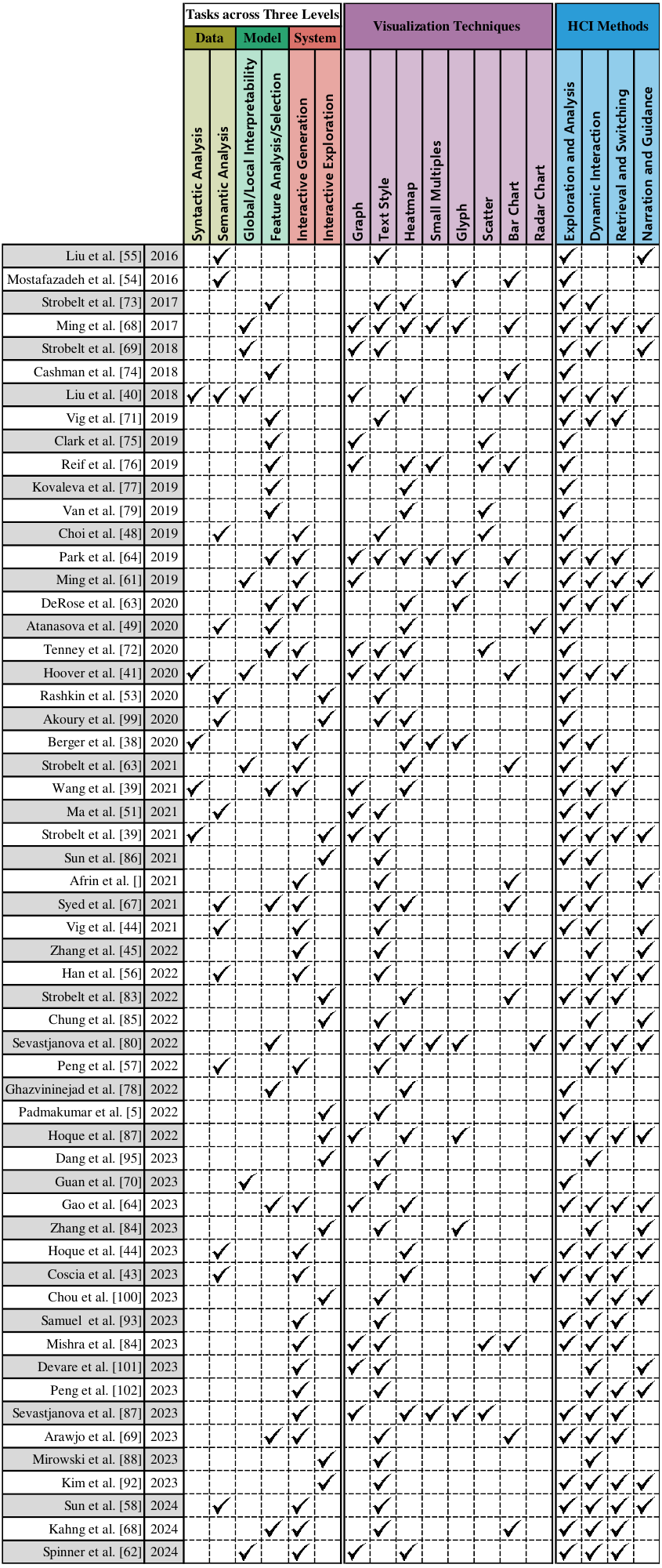}
\caption{
   An overview of the collected papers. We summarize collected papers based on three dimensions. (1) Tasks across three levels: 
    \raisebox{-0.05mm}{\includegraphics[scale=0.01]{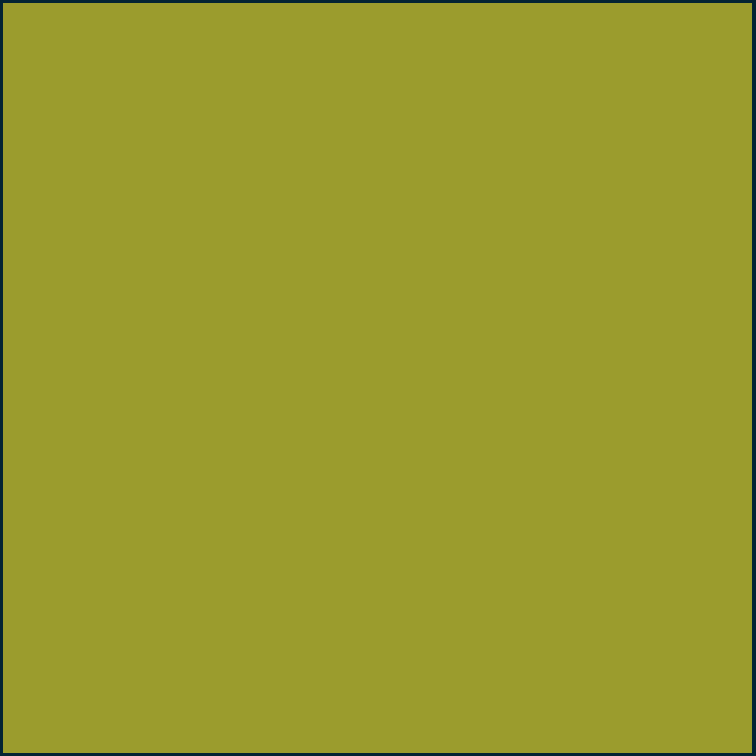}} \textcolor{d1}{Data level} (\setulcolor{d2}\ul{\textit{Syntactic Analysis}} \& \setulcolor{d2}\ul{\textit{Semantic Analysis}}), 
    \raisebox{-0.05mm}{\includegraphics[scale=0.01]{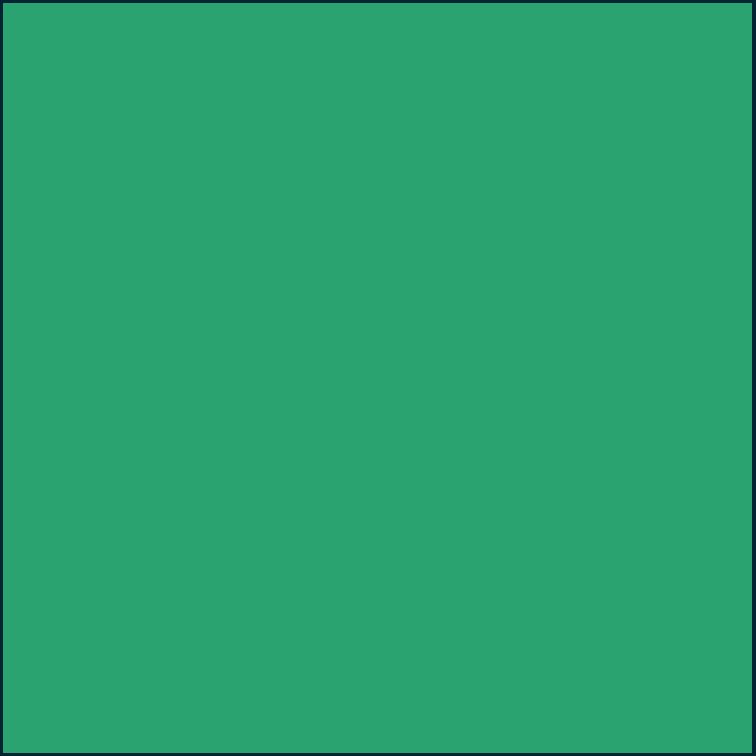}} \textcolor{m1}{Model level} (\setulcolor{m2}\ul{\textit{Global/Local Interpretability}} \& \setulcolor{m2}\ul{\textit{Feature Analysis/Selection}}), and 
    \raisebox{-0.05mm}{\includegraphics[scale=0.01]{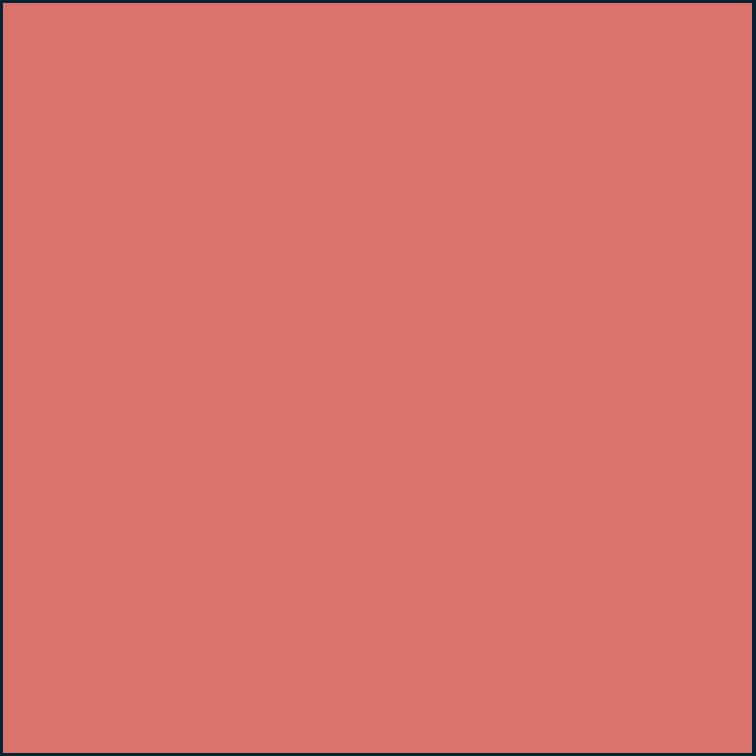}} \textcolor{s1}{System level} (\setulcolor{s2}\ul{\textit{Interactive Generation}} \& \setulcolor{s2}\ul{\textit{Interactive Exploration}}). 
    (2) \raisebox{-0.05mm}{\includegraphics[scale=0.01]{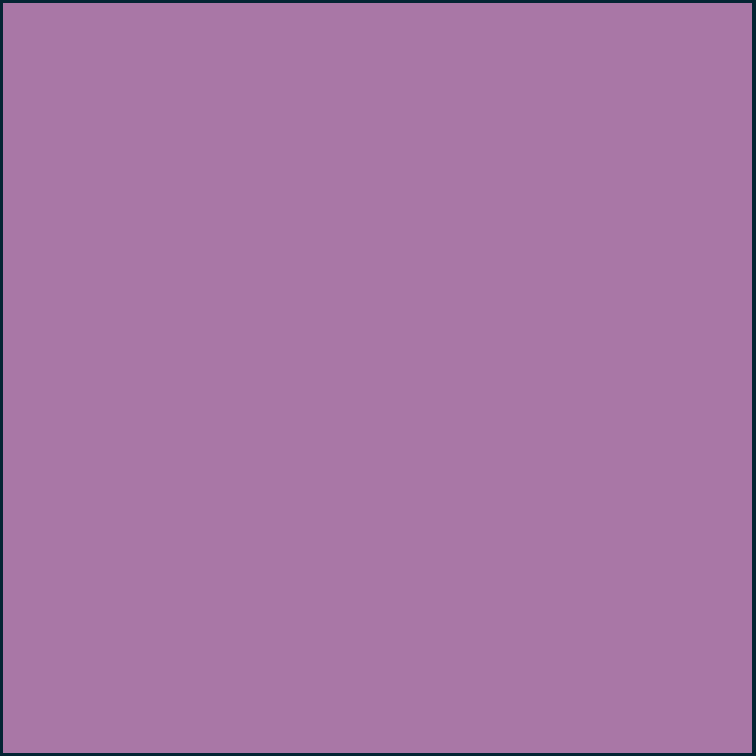}}
\textcolor{v1}{Visualization techniques}, and (3) \raisebox{-0.05mm}{\includegraphics[scale=0.01]{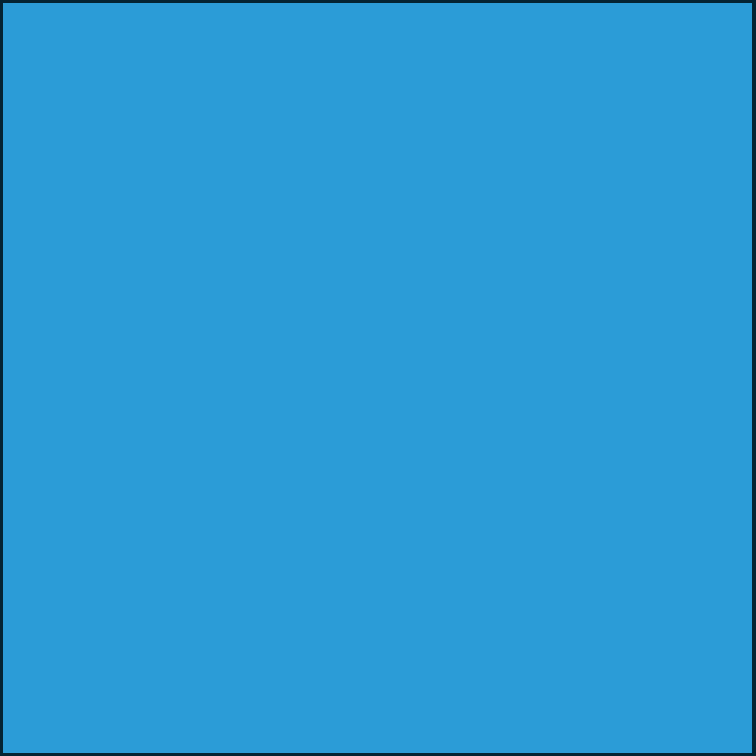}}
\textcolor{h1}{HCI methods}.
    }
\label{fig:Overview}
\end{figure}

\subsection{Taxonomy}
Building upon the high-quality reviews presented in \textit{Section 2}, we propose a taxonomy to systematically categorize HCI methods and visualization techniques employed in text generation tasks based on NLG models. The surveyed papers constitute a comprehensive technical framework encompassing both foundational principles and practical applications. Accordingly, we conduct an in-depth analysis from three distinct perspectives: 
\raisebox{-0.8mm}{\includegraphics[scale=0.02]{figures/FCS-250356-figd1.PNG}} \textcolor{d1}{Data level}, 
\raisebox{-0.8mm}{\includegraphics[scale=0.02]{figures/FCS-250356-figm1.PNG}} \textcolor{m1}{Model level}, and 
\raisebox{-0.8mm}{\includegraphics[scale=0.02]{figures/FCS-250356-figs1.PNG}} \textcolor{s1}{System level}.
Fig.~\ref{fig:Overview} presents an overview of a selection of the collected papers.
Furthermore, we propose two tasks for each level.
Specifically, at the data level, the focus lies on 
\raisebox{-0.8mm}{\includegraphics[scale=0.02]{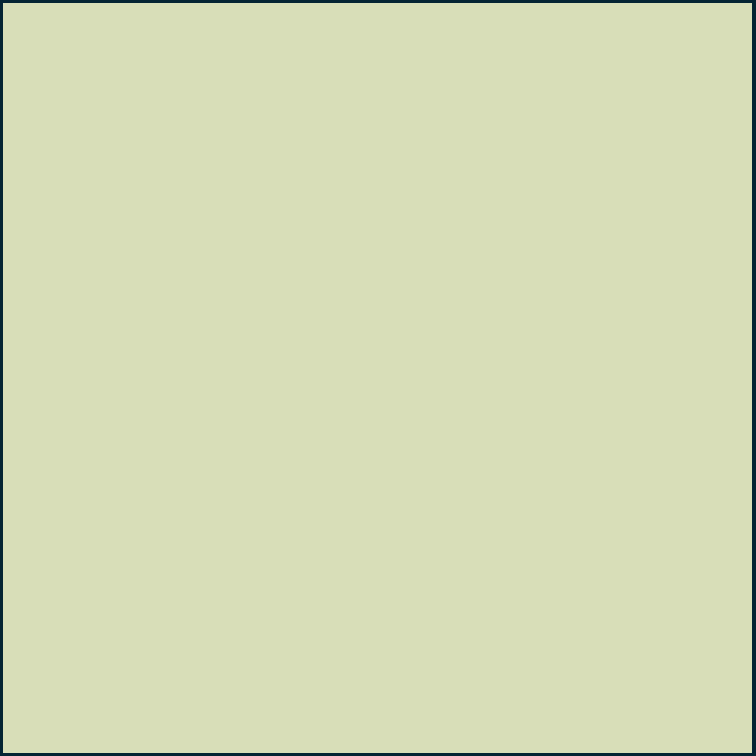}}
\setulcolor{d2}\ul{\textit{Syntactic Analysis}} and \setulcolor{d2}\ul{\textit{Semantic Analysis}}; at the model level, on 
\raisebox{-0.8mm}{\includegraphics[scale=0.02]{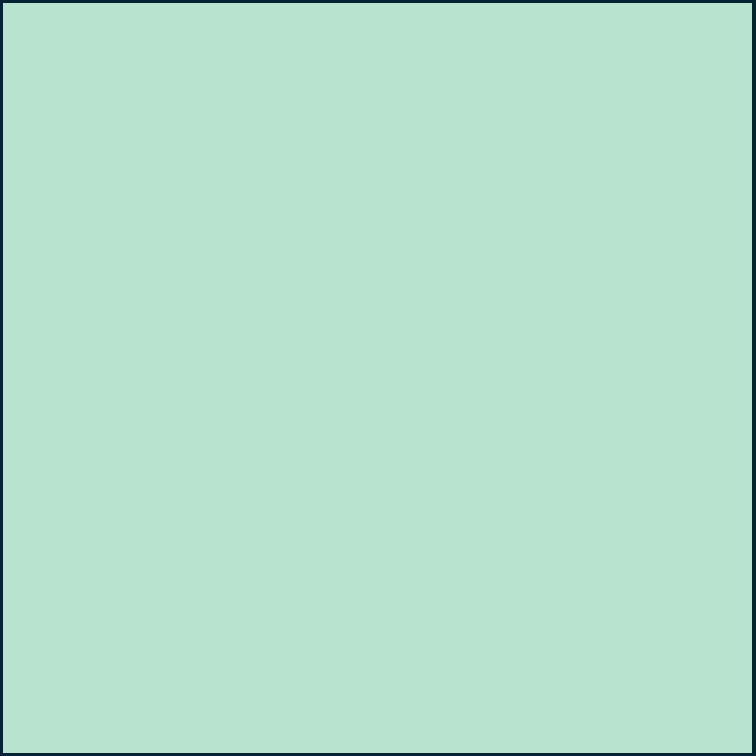}}
\setulcolor{m2}\ul{\textit{Global/Local Interpretability}} and \setulcolor{m2}\ul{\textit{Feature Analysis/Selection}}; and at the system level, on 
\raisebox{-0.8mm}{\includegraphics[scale=0.02]{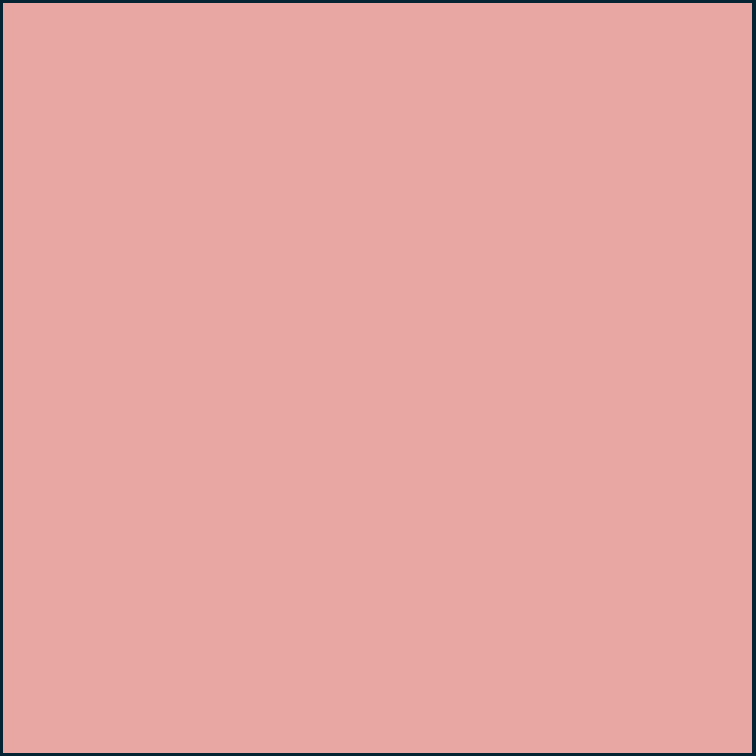}}
\setulcolor{s2}\ul{\textit{Interactive Generation}} and \setulcolor{s2}\ul{\textit{Interactive Exploration}}.

$\bullet$ \textcolor{d1}{\textbf{Data level}}: High-quality data forms the foundation of effective text generation in language models. Data collection, preprocessing, and maintaining linguistic diversity are critical for optimizing model training and improving the quality of generated outputs. In textual data, syntactic structure and semantic content are key. {\setulcolor{d2}\ul{\textit{Syntactic analysis}}}, by identifying and parsing sentence structures, supports the modeling of grammatical relationships. {\setulcolor{d2}\ul{\textit{Semantic analysis}}} enables the interpretation of meaning at the lexical and sentence levels, facilitating the generation of coherent and contextually appropriate text.

$\bullet$ \textcolor{m1}{\textbf{Model level}}: The design and optimization of model architectures are critical to technical implementation, directly influencing a model’s comprehension and generative capabilities, and serving as key drivers of technological advancement. For natural language models, analyzing global and local behaviors, alongside internal feature selection, constitutes an effective strategy for enhancing model efficiency and reliability in real-world applications. {\setulcolor{m2}\ul{\textit{Global interpretability}} facilitates understanding of the model’s overall decision-making process, while {\setulcolor{m2}\ul{\textit{local interpretability}} reveals its response patterns under specific input conditions. {\setulcolor{m2}\ul{\textit{Feature analysis}} supports model optimization by identifying key internal representations, and 
{\setulcolor{m2}\ul{\textit{feature selection}} can reduce model complexity and improve training and inference efficiency.

$\bullet$ \textcolor{s1}{\textbf{System level}}: Effective system integration is crucial not only to the realization of practical applications, but also to the improvement of user interaction and system scalability. As the terminal component of the technological pipeline, the system serves as an interface between the model and end users, facilitating real-time interaction during the text generation process. This interactive framework enables users to refine outputs based on iterative input and feedback, while also allowing for deeper exploration of textual structure and semantics. At the system level, we categorize related works into two functional paradigms: {\setulcolor{s2}\ul{\textit{interactive generation}} and {\setulcolor{s2}\ul{\textit{interactive exploration}}. This classification provides insights into the design of application layer systems that leverage NLG models for content creation and data-driven analysis, thereby guiding future research in building more effective user-facing tools.

We summarize \raisebox{-0.8mm}{\includegraphics[scale=0.02]{figures/FCS-250356-figv1.PNG}}
\textcolor{v1}{visualization techniques} presented in the paper collected.
In the realm of text generation leveraging NLG models, prevalent visualization techniques encompass: \raisebox{-0.8mm}{\includegraphics[scale=0.02]{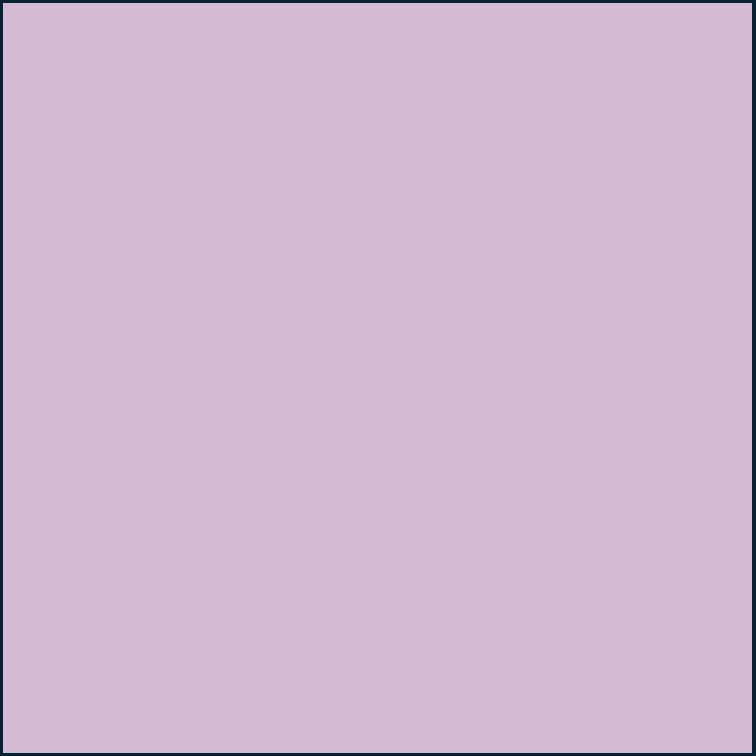}}
{\setulcolor{s2}\ul{\textit{graphs}}, 
{\setulcolor{s2}\ul{\textit{text styles}}, 
{\setulcolor{s2}\ul{\textit{heatmaps}}, 
{\setulcolor{s2}\ul{\textit{small multiples}}, 
{\setulcolor{s2}\ul{\textit{glyphs}}, 
{\setulcolor{s2}\ul{\textit{scatters}}, 
{\setulcolor{s2}\ul{\textit{bar charts}}, 
{\setulcolor{s2}\ul{\textit{radar charts}}.
Graph-based representations play a central role in knowledge modeling and reasoning, while also facilitating context modeling and enabling structured control in generative tasks.
In addition, text styles apply variations in typography and color to highlight key content, thus improving the salience and interpretability of information. 
Heatmaps are a widely adopted visualization technique, commonly used to illustrate attention weights, feature distributions, and generation quality in machine learning models. 
Meanwhile, small multiples present related data views in parallel, enabling intuitive comparison and pattern discovery.
Well-designed glyphs not only enhance readability, but also improve encoding efficiency, allowing high-dimensional information to be effectively embedded and communicated through text. 
Bar charts, with their clarity in displaying numerical comparisons, are frequently used to represent feature importance, output distributions, or performance metrics, which supports a quick evaluation of key factors. 
Finally, radar charts are more suitable for multidimensional performance comparison, particularly in visualizing attention mechanisms and evaluating different modeling strategies.

Additionally, we further propose a classification scheme for \raisebox{-0.8mm}{\includegraphics[scale=0.02]{figures/FCS-250356-figh1.PNG}} \textcolor{h1}{HCI methods}, categorizing them into four types: \raisebox{-0.8mm}{\includegraphics[scale=0.02]{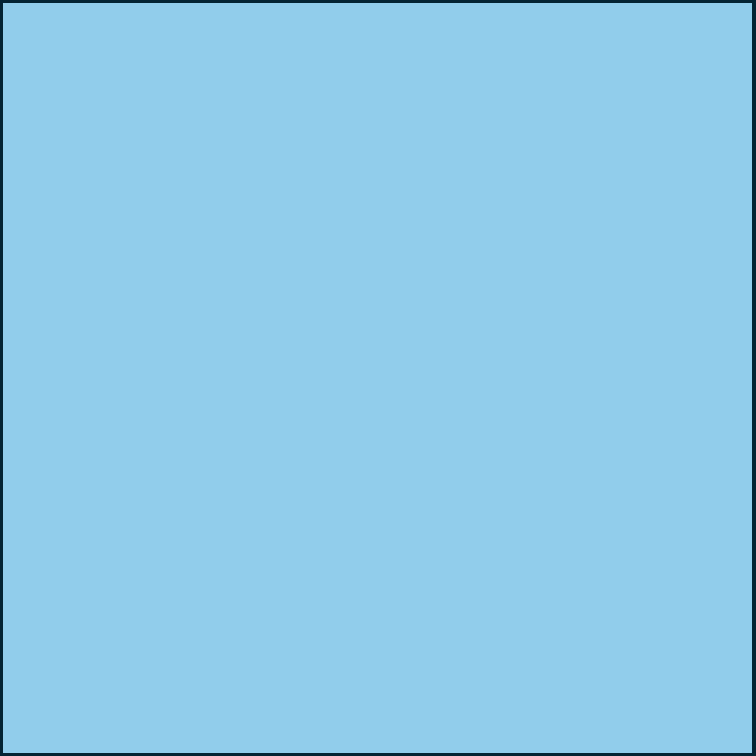}} {\setulcolor{h2}\ul{\textit{Exploration and Analysis (EA)}}, {\setulcolor{h2}\ul{\textit{Dynamic Interaction (DI)}}, {\setulcolor{h2}\ul{\textit{Retrieval and Switching (RS)}}, and 
{\setulcolor{h2}\ul{\textit{Narration and Guidance (NG)}}.

$\bullet$ {\setulcolor{h2}\ul{\textit{EA}} encompasses interactive behaviors such as ``data screening and filtering'', ``multidimensional data exploration'', and ``zooming and panning''. Data screening and filtering enable users to isolate specific data subsets, thereby improving analysis efficiency. Multidimensional data exploration facilitates the examination of complex relationships among multiple variables. Zooming and panning support both macro- and micro-level inspections, allowing users to navigate across different spatial and hierarchical dimensions of the dataset.

$\bullet$ {\setulcolor{h2}\ul{\textit{DI}} involves interactive mechanisms including ``dynamic updates'' and ``linked interactions''. Dynamic updates refer to the real-time synchronization of visual representations in response to data changes or user actions, ensuring timely and accurate information delivery. Linked interactions allow user operations in one view to propagate across multiple coordinated views, thereby supporting integrated and coherent analytical workflows.

$\bullet$ {\setulcolor{h2}\ul{\textit{RS}} includes behaviors such as ``information retrieval'' and ``view switching''. Information retrieval enables users to access supplementary information through interactive elements such as hovering or clicking, without disrupting the primary context. View switching provides flexible transitions between alternative data representations, supports diverse analytical perspectives, and facilitates task-specific exploration.

$\bullet$ {\setulcolor{h2}\ul{\textit{NG}} adopts a storytelling paradigm, embedding visualizations within a narrative structure. This approach improves interpretability by guiding users through data-driven insights in a sequential and context-sensitive manner, thus improving user engagement and comprehension.

\section{Data Level: Generation and Insight}
Data remains a key determinant of the quality of outcomes in the training of the machine learning model~\cite{hsu2019visual}. In the domain of NLP, syntactic structures and semantic information are of paramount importance to researchers and practitioners. The integration of HCI and visual analytics further facilitates users in deriving meaningful insights from the data. Fig.~\ref{fig:datalevel} and Fig.~\ref{fig:datalevel2} illustrate seminal works in this area, showcasing visualization techniques and HCI methods, as summarized in Table~\ref{tab:t1}.

\begin{figure}[htbp]
    \centering
    \includegraphics[width=0.98\linewidth]{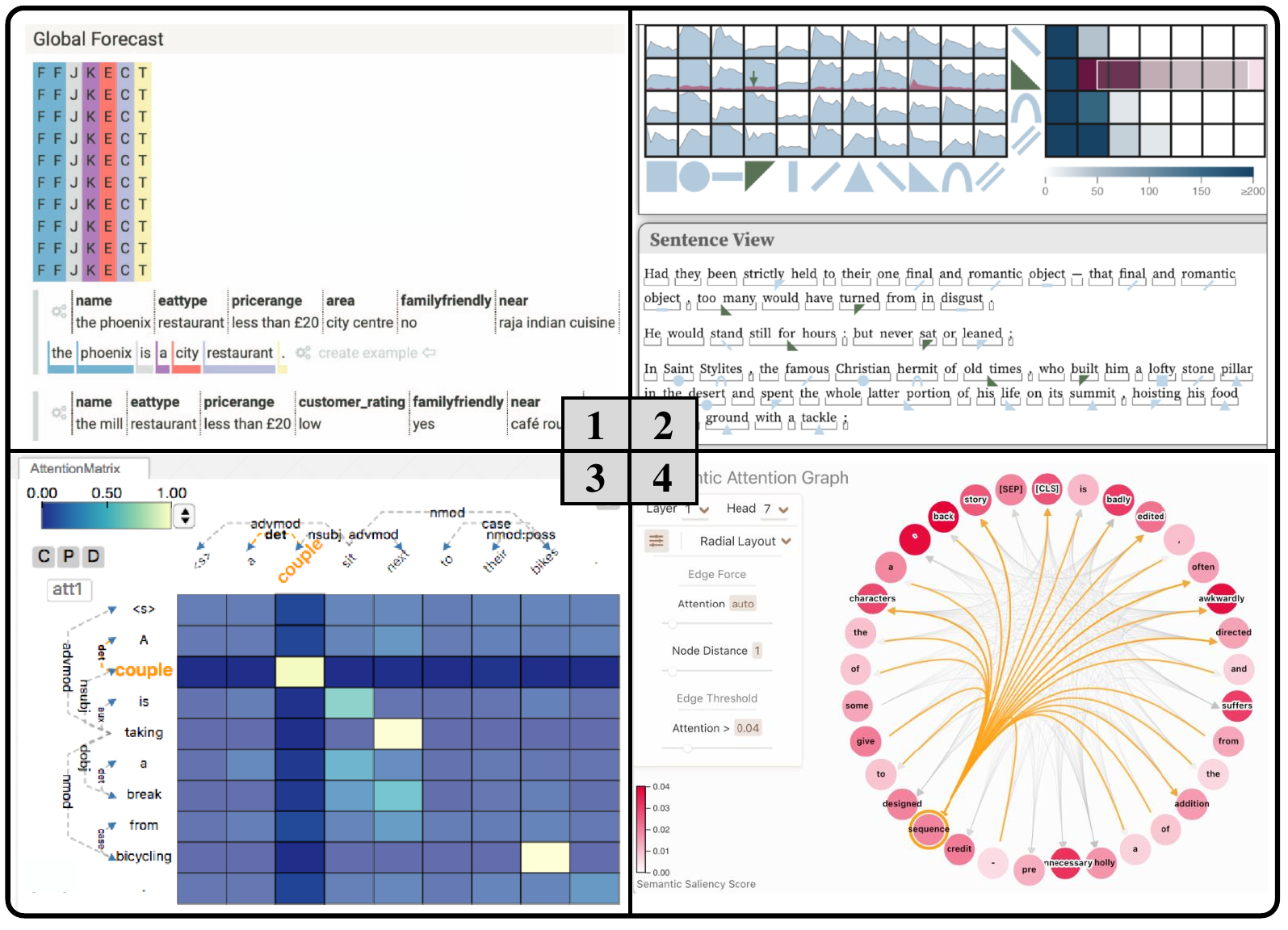}
    \caption{Representative works of syntactic analysis tasks in data level. (1) \textit{GenNI}: An interface for collaborative text generation, which employs well-defined constraints to guide model outputs~\cite{strobelt2021genni}. (2) A method for visual analysis of context embeddings generated by deep neural network-based language models~\cite{berger2020visually}. (3) \textit{Nlize}: An analytical and interpretative tool for natural language inference models~\cite{liu2018nlize}. (4) \textit{Dodrio}: An open-source interactive visualization tool~\cite{wang2021dodrio}.}
    \label{fig:datalevel}
\end{figure}

\begin{table*}[htbp]
\centering
\caption{Visualization techniques and HCI methods in Data Level}
\begin{tabular}{p{4cm}p{6cm}p{3cm}}
\hline
\multicolumn{1}{c}{Representative works}                          & \multicolumn{1}{c}{Visualization techniques} & \multicolumn{1}{c}{HCI methods} \\ \hline
Strobelt et al.~\cite{strobelt2021genni}    & Graph, Text style                            & EA, DI, RS, NG                  \\
Berger et al.~\cite{berger2020visually}     & Heatmap, Glyph, Small multiples              & EA, DI                          \\
Wang et al.~\cite{wang2021dodrio}           & Heatmap, Graph                               & EA, DI, RS                      \\
Liu et al.~\cite{liu2018nlize}              & Scatter, Heatmap, Graph, Bar chart           & EA, DI, RS                      \\
Hoover et al.~\cite{hoover2020exbert}       & Heatmap, Graph, Bar chart, Text style        & EA, DI, RS                      \\
Hoque et al.~\cite{hoque2023portrayal}      & Heatmap                                      & EA, DI, RS, NG                  \\
Coscia et al.~\cite{coscia2023knowledgevis} & Heatmap, Radar chart                         & EA, DI, RS                      \\
Vig et al.~\cite{vig2021summvis}            & Text style                                   & EA, DI, RS                      \\ \hline
\end{tabular}
\label{tab:t1}
\end{table*}

\subsection{Syntactic analysis}
The syntactic structure of natural language is grounded in grammatical rules and linguistic principles, such as part-of-speech (POS) tagging~\cite{strobelt2021genni, berger2020visually} and dependency grammar~\cite{wang2021dodrio, liu2018nlize}. POS tagging is typically formulated as a classification problem and can be visualized using color encoding schemes to distinguish word categories. For example, \textit{GenNi}~\cite{strobelt2021genni} applies color coding to generated tokens, enabling users to quickly assess the compositional structure during global prediction. Shape-based encodings have also been explored to represent categorical distinctions~\cite{berger2020visually}. In contrast, dependency grammar models syntactic relations as directed links between words, often represented as graphs. Systems such as \textit{Dodrio}~\cite{wang2021dodrio} and \textit{Nlize}~\cite{liu2018nlize} incorporate dependency structures with attention mechanisms, allowing for more accurate modeling of both semantic content and syntactic dependencies.

HCI methods for syntactic structure analysis primarily encompass \textit{EA}, \textit{DI}, and \textit{RS}. The objective is to investigate the distribution of syntactic structures within textual data. Although color encoding provides a high-level overview (see Fig.~\ref{fig:datalevel}.1), in-depth analysis often necessitates interactive ``linkage exploration'' within the \textit{DI} component. By representing sentences as word-level vectors and using multidimensional data exploration techniques in \textit{EA}, users can investigate syntactic patterns and evaluate linguistic representations learned by language models (see Fig.~\ref{fig:datalevel}.2). Furthermore, \textit{RS} supports fine-grained inspection by allowing users to hover over words and reveal syntactically related tokens through the ``detail viewing'' mechanism. Given the close association between syntactic structures and attention mechanisms, a unified discussion of attention-based analysis is deferred to \textit{Section 5} for consistency.

\begin{figure}[htbp]
\centering
\includegraphics[width=0.98\linewidth]{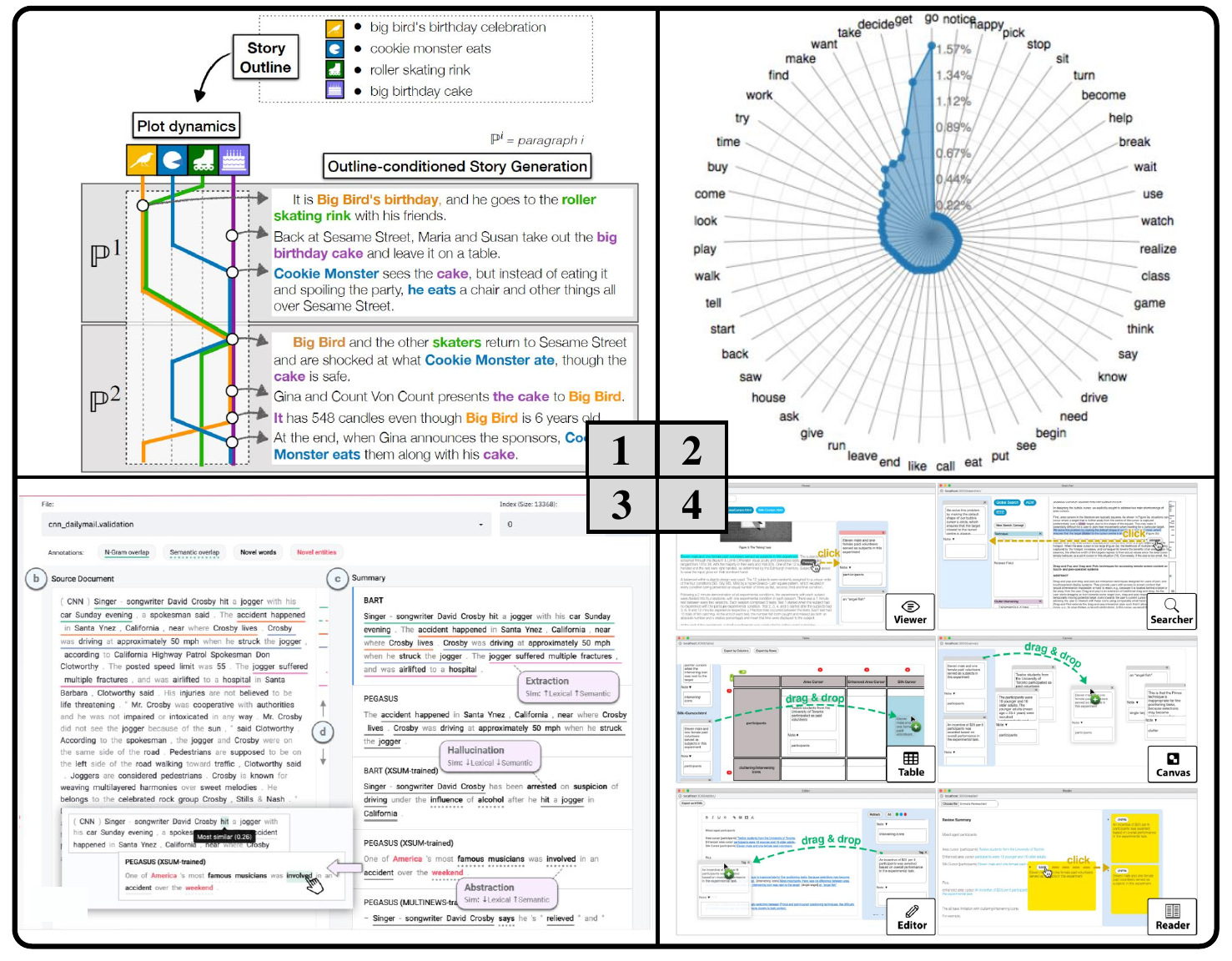}
\caption{Representative works of semantic analysis tasks in data level. (1) \textit{PlotMachines} defines an outline as phrases describing key characters and events~\cite{rashkin2020plotmachines}. (2) A framework can assess corpora and cloze tests to gain a deeper understanding of commonsense narratives~\cite{mostafazadeh2016corpus}. (3) \textit{Summvis} introduces an open-source tool for visualizing abstractive summaries~\cite{vig2021summvis}. (4) \textit{Passages} processes related documents to refine text selection for use, reuse, and sharing across tools~\cite{han2022passages}.}
\label{fig:datalevel2}
\end{figure}

\subsection{Semantic analysis}
Advanced semantic analysis empowers NLG models not only to generate semantically coherent and contextually appropriate text and summaries, but also to perform a wide range of complex natural language processing tasks, including question answering~\cite{gao2021meaningful,zhang2022storybuddy}, dialogue generation~\cite{calderwood2020novelists,biermann2022tool}, and text classification~\cite{choi2019aila,atanasova2020diagnostic}. Here, we focus on text and summary generation tasks. Text generation typically involves producing extended narratives conditioned on specific prompts or semantic associations, while summarization entails the condensation of longer passages into concise representations. In narrative generation, it is common to construct stories following predefined plot trajectories~\cite{li2013story,ma2021eventplus,wang2024e2}. Graph-based representations are often employed to structure narrative elements, facilitating user navigation through characters and events~\cite{rashkin2020plotmachines}, as shown in Fig.~\ref{fig:datalevel2}.1. The categorization of events within texts can be further improved using visualization techniques such as color coding and heat maps~\cite{hoque2023portrayal}. For sentence completion tasks, users benefit from visual indicators that reflect the relevance of candidate words. The bar charts~\cite{mostafazadeh2016corpus} and heatmap~\cite{coscia2023knowledgevis} have proven effective for representing word ranking and alignment. In summary generation, sentence-level semantic attributes such as emotion, topic, or intent can be effectively encoded through controlled variations in text style or formatting~\cite{liu2016automated,vig2021summvis}.

In semantic analysis, HCI methods are employed to support cross-document or cross-paragraph analysis~\cite{han2022passages,peng2022crebot,sun2024reviewflow}. As shown in Fig.~\ref{fig:datalevel2}.4, users can interactively manipulate the sidebars — by zooming and dragging — to refine text selection. By integrating coordinated views and interaction linking, the system enables multiperspective exploration of the same data. Semantic analysis is typically embedded within application systems; for consistency in categorization, these systems are discussed separately in \textit{Section 6}.

\subsection{Discussion}
HCI methods and visualization techniques play a crucial role in both syntactic and semantic analysis. At the data level, visualization techniques are widely employed. To analyze data distribution and magnitude, visualization methods with statistical capabilities, such as heatmaps and bar charts, are frequently utilized. Graphs and radar charts, which capture relationships among sentence components, are also commonly adopted. However, the range of HCI methods applied remains limited; most studies rely primarily on \textit{EA}, \textit{DI}, and \textit{RS}, consistent with the primary focus of the data layer on analytical tasks. Future work may consider integrating more \textit{NG} approaches, enabling data analysis to transition from purely statistical representations to narrative-driven forms, thereby facilitating storytelling as a means of conveying insights.

\section{Model Level: Analysis and Interpretability}
HCI methods and visualization techniques are instrumental in enhancing global/local explainability of NLG models, facilitating deeper insight into model decision-making. Feature analysis and selection further contribute to improving model performance and interpretability. The integration of HCI and visualization approaches enables identification of influential features, thereby supporting model refinement and optimization. Fig.~\ref{fig:Modellevel} and Fig.~\ref{fig:Modellevel2} illustrate representative work in this domain, with corresponding methods summarized in Table~\ref{tab:t2}. These aspects will be discussed in detail below.

\subsection{Global/Local Explainability of NLG models}
Global interpretability aims to characterize the overall behavior of the model, while local interpretability focuses on understanding the model’s behavior and generation mechanisms under specific input conditions. In this work, we concentrate on the input–output interactions of the model, rather than structural modifications to the model architecture. During sentence generation, RNN-based language models transform input data into sequences. The model then predicts subsequent tokens by evaluating semantic similarity with prior inputs, effectively performing a fill-in-the-blank operation. This vocabulary prediction process can be abstracted as a search tree, where each token corresponds to a node. The resulting search space can be visualized using a sankey chart, a method commonly adopted in graph visualization\cite{spinner2024generaitor,ming2017understanding,strobelt2018s}.

\begin{table*}[htbp]
\centering
\caption{Visualization techniques and HCI methods in Data Level}
\begin{tabular}{p{3cm}p{9cm}p{2.5cm}}
\hline
\multicolumn{1}{c}{Representative works}                          & \multicolumn{1}{c}{Visualization techniques} & \multicolumn{1}{c}{HCI methods} \\ \hline
Ming et al.~\cite{ming2019protosteer}           & Graph, Glyph, Bar chart  & EA, DI, RS, NG                  \\
Spinner et al.~\cite{spinner2024generaitor}     & Graph, Heatmap        & EA, DI, RS                      \\
Strobelt et al.~\cite{strobelt2021lmdiff}       & Heatmap, Bar chart    & EA, RS                          \\
Park et al.~\cite{park2019sanvis}               & Heatmap, Graph, Glyph, Bar chart, Small multiple, Text style & EA, DI, RS                      \\
DeRose et al.~\cite{derose2020attention}        & Glyph, Heatmap        & EA, DI, RS                      \\
Gao et al.~\cite{gao2023transforlearn}          & Heatmap, Graph        & EA, DI, RS, NG                  \\
Syed et al.~\cite{syed2021summary}      & Heatmap, Text style, Bar chart                               & EA, DI, RS                      \\
Kahng et al.~\cite{kahng2024llm}                & Text style, Bar chart & EA, DI, RS                      \\
Arawjo et al.~\cite{arawjo2023chainforge}       & Text style, Bar chart & EA, DI, RS                      \\ \hline
\end{tabular}
\label{tab:t2}
\end{table*}

\textit{ProtoSteer}~\cite{ming2019protosteer} introduces an interactive visualization approach for interpretable and controllable sequence modeling. By integrating HCI methods such as detail viewing and view switching (\textit{IV}), along with linked interactions (\textit{DI}), the system employs graph, glyph, and bar chart to support modeling scenarios. \textit{LMDIFF}~\cite{strobelt2021lmdiff} enables comparative analysis of token-level differences across identical prompts by leveraging \textit{DI} and \textit{RS} interactions. It combines heatmaps and bar charts to evaluate performance variations across NLG models. \textit{ERIC}~\cite{guan2023generating} focuses on guiding text generation via learning dynamic and discrete entity states within a contrastive learning framework. It incorporates text style features to retrieve generated sentences that resemble training samples in content.

\begin{figure}[htbp]
    \centering
    \includegraphics[width=0.98\linewidth]{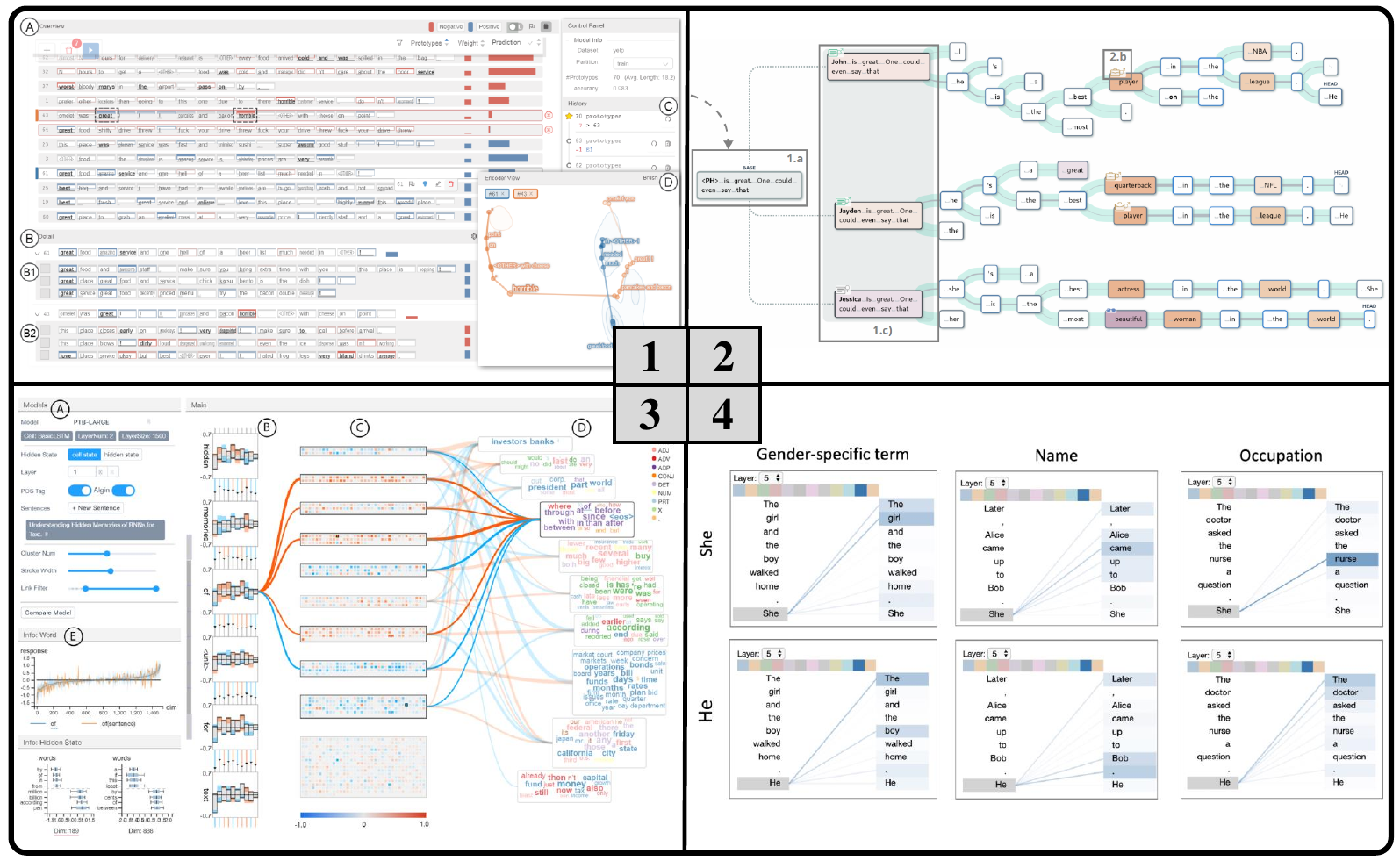}
    \caption{Representative works of global and local explainability tasks in model level. (1) \textit{ProtoSteer} introduces an interactive visualization approach for interpretable and manipulable deep sequence modeling~\cite{ming2019protosteer}. (2) \textit{generAIto} employs a visual analytics technique that enhances search trees with task-specific widgets~\cite{spinner2024generaitor}. (3) \textit{RNNVis} is used for understanding and comparing the performance of RNN models in NLP tasks~\cite{ming2017understanding}. (4) involves multiscale visualization of attention in the Transformer model for visual analysis~\cite{vig2019multiscale}.}
    \label{fig:Modellevel}
\end{figure}

\subsection{Feature Analysis and Selection Techniques}
Analyzing model characteristics and adjusting weights can improve model interpretability. Feature selection and model compression further enhance the efficiency of both training and inference~\cite{tenney2020language}. This discussion focuses on the internal representation and structural modification of models. It is well recognized that different architectures are grounded in distinct computational paradigms. LSTM, a specific form of RNN, is widely adopted for sequential data processing. In contrast, Transformer-based models evolve through successive refinements of the attention mechanism. Regardless of the underlying architecture, feature-level performance remains a critical concern for model optimization. For RNNs, analyzing token-level performance within sequences is essential. \textit{LSTMVIS}~\cite{strobelt2017lstmvis} employs heatmaps and typographic cues to visualize token behaviors, while \textit{RNNbow}~\cite{cashman2018rnnbow} uses bar charts. 

In Transformer-based models, attention mechanisms are the primary focus of interpretability research. Graphs are employed to analyze how attention heads at different layers capture diverse dependency relationships~\cite{clark2019does,vig2019multiscale}. Heatmaps illustrate the intensity of these relationships~\cite{reif2019visualizing,kovaleva2019revealing,ghazvininejad2022discourse}, while scatter plots reveal the information most relevant to question answering across layers~\cite{van2019does}. To manage the increased complexity introduced by multi-headed self-attention networks, \textit{SANVis}~\cite{park2019sanvis} supports data filtering and sorting via sliders and buttons (\textit{EA}), and enables detailed inspection of individual attention heads through mouse interactions and view switching (\textit{RS}). Furthermore, linked interactions within the visual interface (\textit{DI}) improve the effectiveness of data exploration. \textit{LMFingerprints}~\cite{sevastjanova2022lmfingerprints} identifies the layers involved in specific analytical tasks by tracking the propagation of contextual information throughout the model.

In the domain of visualization techniques, \textit{Attention Flows}~\cite{derose2020attention} differs from \textit{SANVis} by employing not only heatmaps to represent the weights of attention heads, but also glyph-based representations to enhance the interpretability of attention dynamics across word sequences and attention layers. When introducing the Transformer model, \textit{TransforLearn}~\cite{gao2023transforlearn}, as shown in Fig.~\ref{fig:Modellevel2}.3, presents the model's processing pipeline in a form accessible to non-expert users. \textit{Summary Explorer}~\cite{syed2021summary} enables comparative analysis of summarization models by allowing users to select specific features and contrast their outputs. In contrast, \textit{LLM Comparator}~\cite{kahng2024llm} avoids reliance on performance scores; instead, it employs an interactive workflow to help users understand when and why a model outperforms or underperforms a baseline, and how the response quality varies between models. \textit{ChainForge}~\cite{arawjo2023chainforge}, an open-source visualization toolkit, offers a graphical interface for comparative analysis of model outputs and prompt variations.

\begin{figure*}[htbp]
\centering
\includegraphics[width=0.98\linewidth]{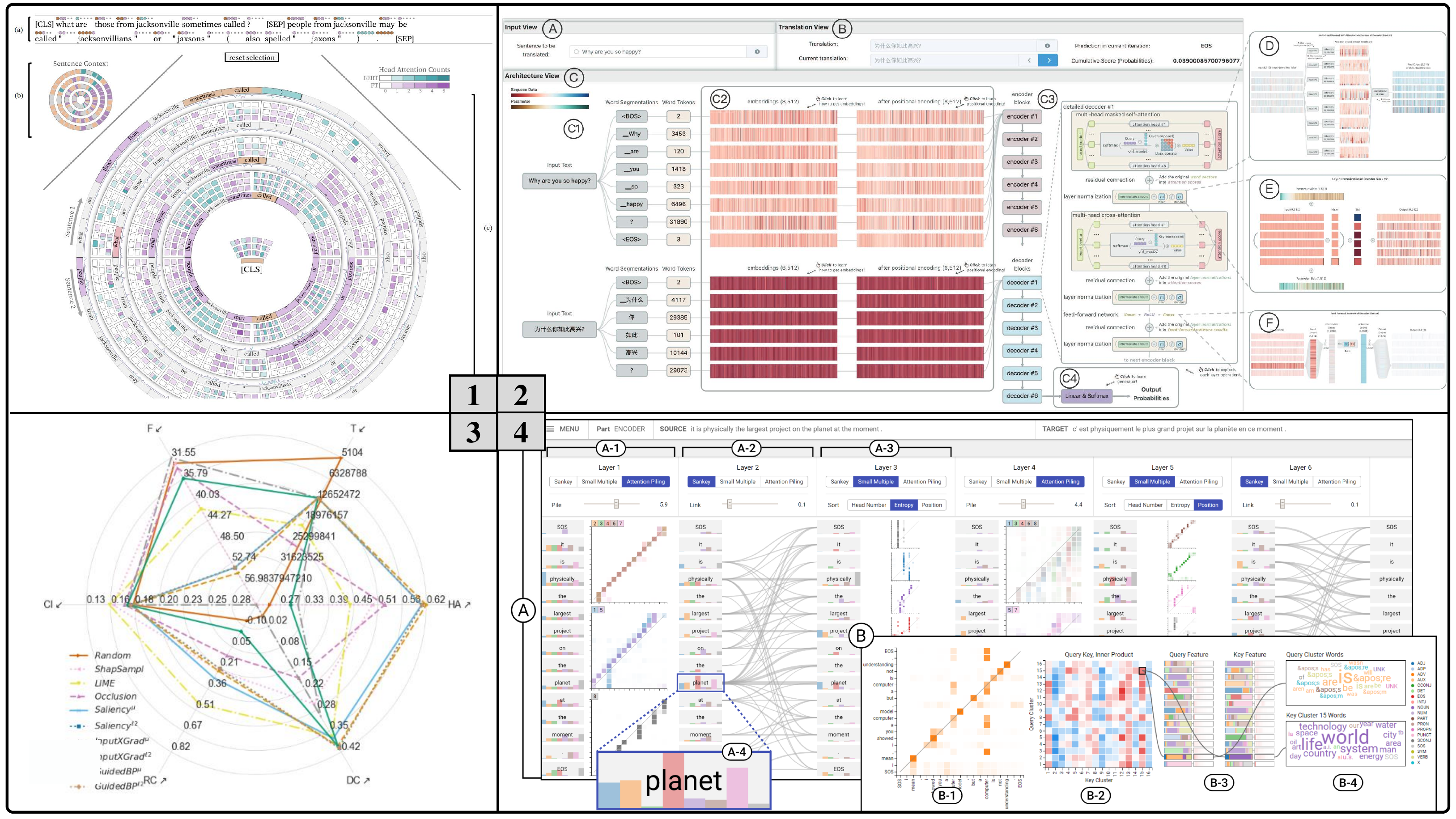}
  \caption{Representative works of feature analysis and feature selection tasks in model level. (1) \textit{Attention flows} employ attention mechanisms to study various sentence comprehension tasks~\cite{derose2020attention}. (2) \textit{TransforLearn} is an interactive visual tutorial that helps beginners understand the Transformer~\cite{gao2023transforlearn}. (3) proposes a comprehensive list of diagnostic attributes to evaluate interpretability techniques from various perspectives~\cite{atanasova2020diagnostic}. (4)\textit{ SANVis} enables deep understanding of multi-head self-attention at multiple levels~\cite{park2019sanvis}.}
\label{fig:Modellevel2}
\end{figure*}

\subsection{Discussion}
In model development, improving the interpretability of NLG models and optimizing model parameters are central research objectives. Analyzing the relationships among model layers, the influence of hierarchical structures on outputs, and the characteristics of generated text is a key approach to uncovering the internal mechanisms of NLG models. At the model level, visualization techniques such as heatmaps, graphs, and text style are frequently used. These methods support the representation of layer-wise weights (heatmaps), inter-layer dependencies (graphs), and the impact of individual layers on generation outcomes (text styles). In the context of HCI, \textit{EA}, \textit{RS}, and \textit{DI} constitute the three primary interaction strategies. For non-expert users, \textit{NG} also plays a critical role by enabling narrative-driven interaction. \textit{NG} helps illustrate the effects of model behavior and parameter changes, thereby facilitating user understanding of NLG models.

\section{System Level: Interaction and Exploration}
Under the system-level perspective, HCI methods in interactive text generation systems support intuitive language-based interactions, improving user engagement and control. Concurrently, visualization techniques enable effective exploration of NLG model outputs by rendering complex data patterns and model behaviors comprehensible. These approaches facilitate deeper insight into generation processes and support informed interpretation of model decisions.

\begin{table*}[htbp]
\centering
\caption{Visualization techniques and HCI methods in System Level}
\begin{tabular}{p{3.5cm}p{7cm}p{3cm}}
\hline
\multicolumn{1}{c}{Representative works}                          & \multicolumn{1}{c}{Visualization techniques} & \multicolumn{1}{c}{HCI methods} \\ \hline
Strobelt et al.~\cite{strobelt2022interactive} & Heatmap, Bar chart                             & EA, DI, RS                      \\
Zhang et al.~\cite{zhang2023visar}             & Glyph, Text style                              & DI, NG                          \\
Chung et al.~\cite{chung2022talebrush}         & Text style                                     & DI, NG                          \\
Mishra et al.~\cite{mishra2023promptaid}       & Scatter, Graph, Bar chart, Text style          & EA, DI, RS                      \\
Sevastjanova et al.~\cite{sevastjanova2023visual}      & Scatter, Graph, Glyph, Heatmap, Small Multiple & EA, DI, RS                      \\ 
Hui et al.~\cite{hui2023lettersmith}      & Text style & DI, NG                      \\ 
Afrin et al.~\cite{afrin2021effective}       & Text style           & DI, NG                      \\
\hline
\end{tabular}
\label{tab:t3}
\end{table*}

\begin{figure}[htbp]
    \centering
    \includegraphics[width=0.98\linewidth]{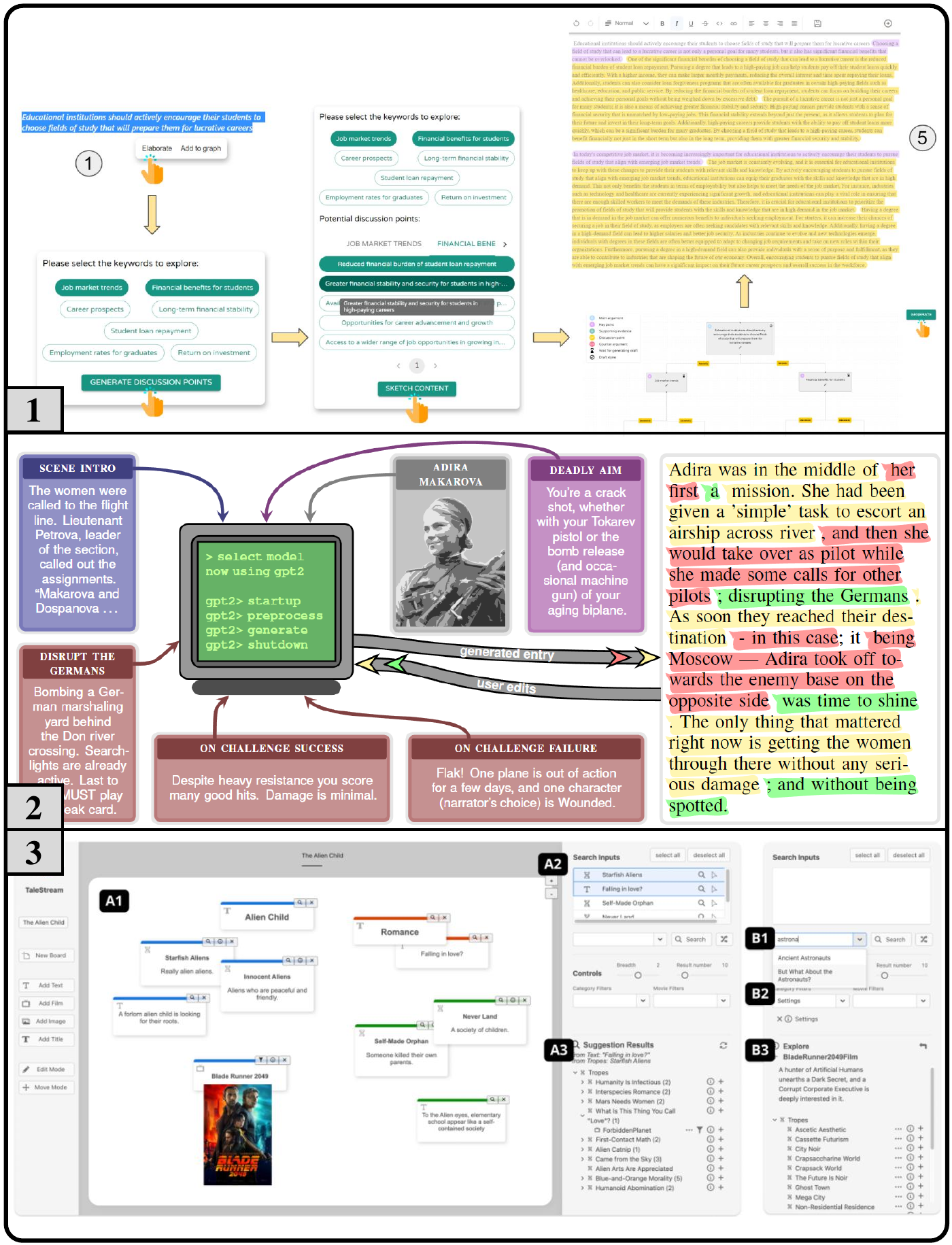}
    \caption{Representative works of interactive generation tasks in system level. (1) \textit{VISAR} is an AI writing assistant that helps authors brainstorm and refine hierarchical writing goals~\cite{zhang2023visar}. (2) \textit{STORIUM} is a dataset and evaluation platform for testing fine-tuned language models~\cite{akoury2020storium}. (3) \textit{TaleStream} is a system that supports story writing with inspiring materials~\cite{chou2023talestream}.}
    \label{fig:Systemlevel1}
\end{figure}

\subsection{Interactive Generation Systems}
An interactive text generation system refers to a system that generates text through real-time interaction with users. They utilize NLG models to understand and generate text. Such systems are capable of producing timely responses based on user input, such as answering questions~\cite{strobelt2022interactive}, providing information~\cite{sun2021iga,hoque2022dramatvis}. 

\begin{figure*}[htbp]
    \centering
    \includegraphics[width=0.98\linewidth]{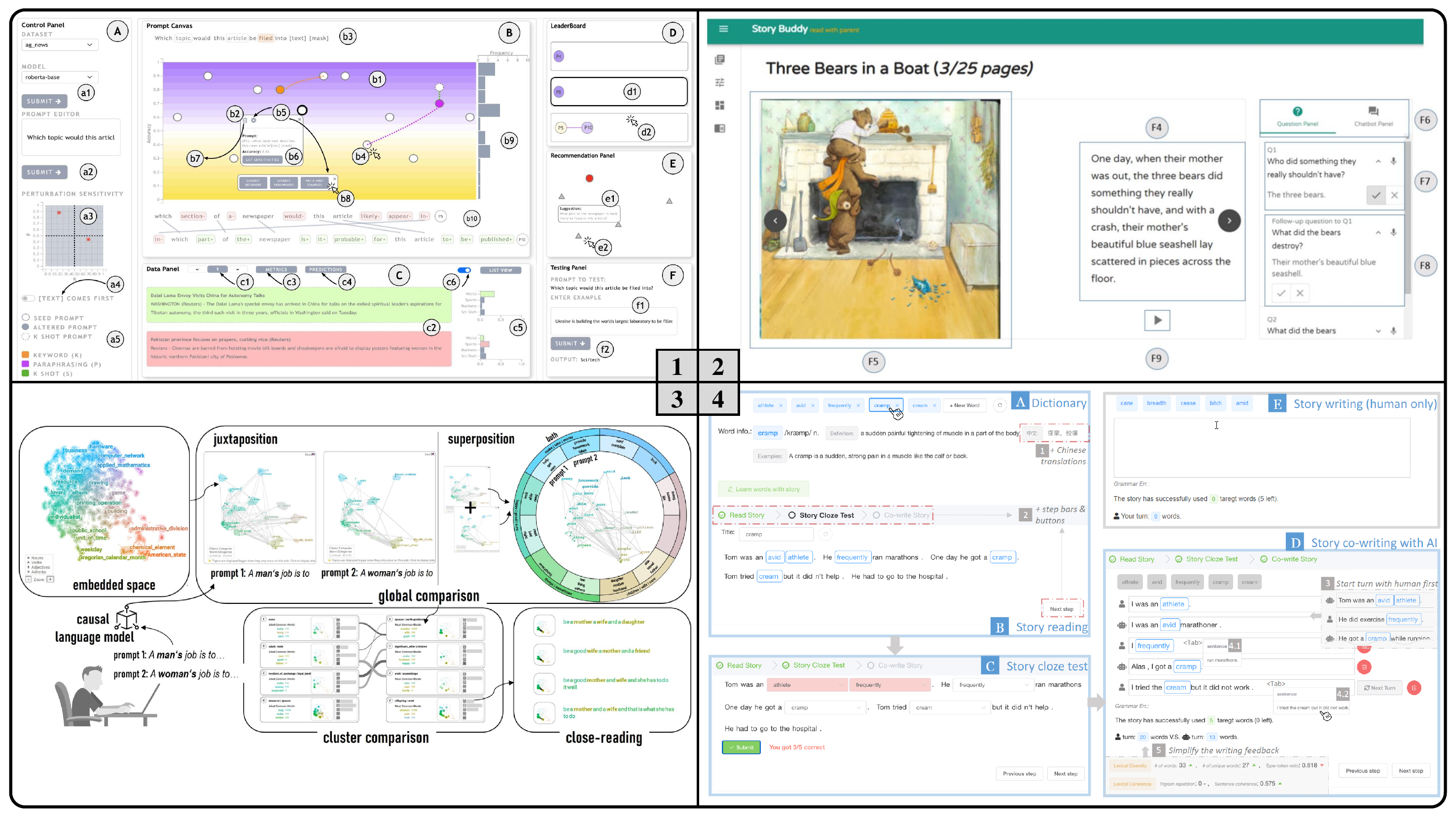}
    \vspace{-0.5em}
    \caption{Representative works of interactive exploration tasks in system level. (1) \textit{PromptAid} is a visual tool for creating, refining, and testing prompts through exploration and interaction~\cite{mishra2023promptaid}. (2) \textit{StoryBuddy} is an AI-supported system for parents to create interactive storytelling experiences~\cite{zhang2022storybuddy}. (3) A visual analysis approach enables users to set initial prompts and group generated text for exploratory analysis of outputs from causal language models~\cite{sevastjanova2023visual}. (4) \textit{Storyfier} uses text generation models to help learners read and write stories with target words, supported by adaptive AI assistance~\cite{peng2023storyfier}.}
    \label{fig:Systemlevel2}
    \vspace{-1.5em}
\end{figure*}

The predominant interaction paradigm relies on dialogue, where users convey their intentions to the model through textual inputs, referred to as prompts~\cite{mirowski2023co}. Despite its ubiquity, this prompt-based interaction incurs substantial communication overhead~\cite{webson2022prompt}. To improve interaction efficiency, recent research focuses on analyzing user-model interaction patterns~\cite{cheng2022mapping,gero2023social}. Non-expert users can also harness the capabilities of language models by systematically refining prompts using principled methodologies~\cite{strobelt2022interactive,kim2023evallm,suvcik2023prompterator}.
\textit{Promptsource}~\cite{bach2022promptsource} supports prompt creation, sharing, and reuse. Dang et al.~\cite{dang2023choice} conducted an analysis of user requirements and perceptions related to system prompts. \textit{IPL}~\cite{jin2023instance}, an instance-aware prompt learning approach, enables the model to learn individualized prompts for each input instance. Beyond user-level prompt engineering, generative interaction paradigms at the model level have emerged. \textit{PromptChainer}~\cite{wu2022promptchainer} and \textit{AI Chains}~\cite{wu2022ai} introduce a compositional framework for handling complex or multi-step tasks, wherein multiple LLMs are sequentially invoked to iteratively refine outputs.

For long-form content generation, prompt control becomes more granular. As illustrated in Fig.~\ref{fig:Systemlevel1}.1, \textit{VISAR}~\cite{zhang2023visar} allows users to dynamically update generated content by specifying high-level paragraph themes. \textit{STORIUM}~\cite{akoury2020storium} enables interactive story construction through a game-like interface. \textit{TaleBrush}~\cite{chung2022talebrush} redefines story authoring by shifting from textual prompts to visual sketching, where narrative elements are mapped to stroke characteristics such as line thickness and drawing duration.

\subsection{Interactive Exploration with NLG models}
Interactive exploration enables real-time interaction between users and models through customized prompts, facilitating the generation of text that satisfies specific requirements~\cite{coscia2023knowledgevis,mishra2023promptaid}. Users can iteratively refine their inputs based on preliminary outputs, and the model correspondingly adjusts its responses, thereby improving both the quality and relevance of the generated content. This process also promotes greater user engagement and satisfaction. Furthermore, interactive exploration supports multi-turn dialogue, allowing for the progressive refinement of queries and responses, which results in more accurate and personalized text generation.

\textit{KnowledgeVIS}~\cite{coscia2023knowledgevis} is a human-in-the-loop visual analytics system that leverages fill-in-the-blank prompts to interpret language models. By interactively comparing sentence-level predictions, it reveals learned associative patterns. 
\textit{PromptAid}~\cite{mishra2023promptaid} facilitates interactive prompt engineering through exploration, perturbation, evaluation, and iteration. It employs coordinated visualizations to enhance prompts via keyword and paraphrase perturbations, as well as through the selection of optimal in-context few-shot examples. 
\textit{SAGEViz}~\cite{devare2023sageviz} integrates NLG models with human oversight to support the creation, expansion, and visualization of event patterns, enabling collaborative and seamless generation and modification of event sequences.

\subsection{Discussion}
At the system level, research primarily investigates methods to improve the efficiency of user interaction with NLG models for both interactive text generation and exploratory tasks. While the overarching design objectives remain consistent, the visualization techniques employed exhibit considerable variation. Despite this, their application tends to follow a relatively uniform pattern. With respect to HCI methods, \textit{DI} receives particular emphasis in facilitating system-user interaction, while \textit{EA}, \textit{RS}, and \textit{NG} are also comparatively balanced. It is hypothesized that when NLG models are developed as general-purpose tools, the selection of visualization techniques and HCI methods becomes more comprehensive, thereby contributing to reduced user adoption barriers.

\section{Conclusion}
In this paper, we conduct a survey that systematically reveals the roles and limitations of HCI methods and visualization techniques throughout the entire process, which from data and models to system applications.
We have found that the integration of HCI methods and visualization techniques has transformed the traditional, tedious approach of merely viewing various numerical data and programming interfaces.

\textbf{Advantages}.
HCI methods improve user experience and simplify interaction with systems, whether dialog-based or peripheral-based. Timely system feedback facilitates rapid user adjustments to generated content, enhancing both accuracy and quality. Additionally, user feedback enables model adaptation and performance optimization. Users can further refine model behavior by tuning parameters or supplying additional contextual information for customized generation.
During content generation, visualization techniques present the underlying processes and results via charts, dashboards, and other visual representations. These visualizations enhance user comprehension and trust by elucidating the model's decision-making mechanisms and generation logic. They also assist in identifying patterns, anomalies, and potential areas for improvement within the generation model.

\textbf{Limitations}.
The design and implementation of efficient HCI systems entail complex engineering processes and substantial resource investment. In particular, effective natural language generation and user interface design demand multidisciplinary collaboration. Despite efforts to ensure intuitiveness, users typically require training and practice to utilize advanced functionalities and customized operations proficiently. These systems depend heavily on user input and feedback for content generation and optimization. However, excessive reliance on user feedback can lead to adaptation issues when encountering new users or unfamiliar contexts.
The development of visualization tools that expose the internal mechanisms of models often incurs significant computational costs and requires specialized expertise. For NLG models, implementing real-time visualizations remains challenging, potentially degrading user experience. Moreover, complex visualizations may hinder interpretability for non-expert users, while excessive informational density can cause cognitive overload, ultimately reducing system usability.

\textbf{Opportunities}.
Incorporating user feedback on model performance, including satisfaction metrics and improvement suggestions, facilitates iterative training and optimization. Advances in HCI techniques enable novel applications of LLMs in domains such as online education and interactive media. Enhanced interaction methods further lower the barrier for non-technical users to engage with LLMs for text generation, thereby promoting broader adoption and acceptance.
Visualization techniques improve the interpretability of model decision-making processes and enhance user trust in AI systems. They also foster user engagement and encourage exploratory interaction, contributing to a deeper understanding of model behavior. The integration of visualization in text generation models can attract interdisciplinary collaboration across computer science, design, and cognitive science, fostering innovation in both research and application.

\begin{acknowledgement}
National Key Research and Development Program of China (2022YFB3104800), National Natural Science Foundation of China (62422607, 62372411, 62432014), Zhejiang Provincial Natural Science Foundation of China (LR23F020003). Guodao Sun is the corresponding author.
\end{acknowledgement}

\begin{competinginterest}
The authors declare that they have no competing interests or financial conflicts to disclose.
\end{competinginterest}




\bibliographystyle{fcs}
\bibliography{ref}

\end{document}